\documentclass[11pt]{article}

\usepackage{amsmath}
\usepackage{amsfonts}
\usepackage{amssymb}
\usepackage{graphicx}
\usepackage{supertabular}
\usepackage{setspace}
\usepackage{xcolor}
\usepackage{mathptmx}

\RequirePackage{cite}%
\renewcommand{\citeleft}{\bgroup\normalfont[}%
\renewcommand{\citeright}{]\egroup}%

\newcommand{\n}{\noindent}
\newcommand{\tr}{{\rm tr\,}}

\begin{document}

\title{\leftline{\footnotesize doi:10.1088/0264-9381/27/1/015012}
\leftline{\footnotesize M. Azreg-A\"{\i}nou, \textit{Class. Quantum Grav.} \textbf{27} (2010) 015012 (16pp)}
\vskip0.5cm
Developed Adomian method for quadratic Kaluza-Klein relativity}
\author{Mustapha Azreg-A\"{\i}nou
\\Ba\c{s}kent University, Department of Mathematics, Ba\u{g}l\i ca Campus, Ankara, Turkey}
\date{}

\maketitle

\begin{abstract}
We develop and modify the Adomian decomposition method (ADecM) to work for a new type of nonlinear matrix differential equations (MDE's) which arise in general relativity (GR) and possibly in other applications. The approach consists in modifying both the ADecM linear operator with highest order derivative and ADecM polynomials. We specialize in the case of a 4$\times$4 nonlinear MDE along with a scalar one describing stationary cylindrically symmetric metrics in quadratic 5-dimensional GR, derive some of their properties using ADecM and construct the \textit{most general unique power series solutions\/}. However, because of the constraint imposed on the MDE by the scalar one, the series solutions terminate in closed forms exhausting all possible solutions.

\vspace{3mm}

\n {\footnotesize\textbf{AMS Classification.} 34G20, 49M27, 15A24, 15A30}

\n {\footnotesize\textbf{Keywords.} Nonlinear matrix differential equations, Decomposition methods, Matrix equations and identities, Cosmic strings in Kaluza-Klein theory}

\vspace{-3mm} \n \line(1,0){431} 
\end{abstract}


\section{Introduction}
Matrix differential equations (MDE's), which arise in many
applications such as biomathematics, economics, electronics,
physics, stability analysis and statistics \cite{Campbell,Lewis},
are increasingly interesting fields of applied mathematics and
integral methods. In 5-dimensional general relativity (5DGR) and
generally in $d$-dimensional general relativity ($d$DGR) with $d>4$,
linear and nonlinear 4$\times$4 or 3$\times$3 MDE's arise
naturally in the 4~+~($d-4$) or 3~+~($d-3$) dimensional
reductions, respectively \cite{Azreg,AzregN,Maison,AzregC}. Applications
of MDE's to 4DGR are also known in the literature~\cite{Clement,Wild}.

The kind of MDE's we deal with in GR are different from the
Riccati, Sylvester or Lyapunov MDE's encountered
in many other applications. In the latter MDE's, the coefficients
of the unknown $N\times N$ matrix, denoted by $\chi$, are given
matrices; while in the former case, these same coefficients are
unknown scalar functions depending explicitly on the invariants
of $\chi$ and their derivatives.

In order to introduce the MDE's encountered in GR we first define
an abstract general form of them. Letting $\ell_i\equiv \tr (\chi^i)$,
$1\leq i\leq N-1$, and $k \equiv \det \chi$, we consider the
following first order nonlinear MDE
\begin{equation}\label{1}
    \sum_{j=1}^{N-1}\,G_j(\ell_i,k,\ell_{i,\rho},k_{i,\rho})(\chi^j)_{,\rho} + \sum_{j=0}^{N-1}\,H_j(\ell_i,k,\ell_{i,\rho},k_{i,\rho})\,\chi^j =0\,,
\end{equation}
where ``$,\rho$" denotes the derivative $\text{d}/\text{d}\rho$ with respect to the real
variable $\rho$ and $G_j$, $H_j$ \& $H_0$ ($1\leq j\leq N-1$) are given
explicit scalar functions of the unknown invariants of $\chi$ and their first order
derivatives. In equation~\eqref{1}, $\chi^0=I$ denotes the $N\times N$ identity
matrix and will be omitted throughout this paper. Notice that the powers of $\chi$ higher than $N-1$
and their traces do not appear in~\eqref{1} for these can be reduced
using the Cayley-Hamilton equation~\cite{Ox}.

Introducing two arbitrary functions $(G_0,T)$ and a matrix $Q$ defined by
\begin{equation}\label{2}
   Q \equiv \sum_{j=0}^{N-1}\,G_j\,\chi^j\,,
\end{equation}
we bring equation~\eqref{1} to the form
\begin{equation}\label{3}
   Q_{,\rho} + TQ + \sum_{j=0}^{N-1}\,R_j\,\chi^j = 0\,,
\end{equation}
where $R_j \equiv H_j - G_{j,\rho} -TG_j$ for $0\leq j\leq N-1$.

In many applications the extra functions $(G_0,T)$ depend, or are
chosen to depend, linearly on $\ell_1 (=\tr \chi)$. Generally speaking,
equation~\eqref{2} can be inverted to express
$\chi$ as a polynomial in $Q$ then substituted in~\eqref{3}
to bring the latter into a form where $Q$ appears as the main variable.
However, when the size of $\chi$ is big such an inversion is not suitable;
secondly, the method of resolution we are applying requires equation~\eqref{3}
to be kept in that form where $\chi$ is the main variable.

The aim of the paper is not to discuss in detail the properties of equation~\eqref{3}
or equation~\eqref{1}, although some of them are formulated in
section~\ref{for} and proven in section~\ref{est}. Rather, one of the purposes of this work
is to develop and apply the Adomian decomposition method (ADecM)~\cite{ad} to
solve this type of nonlinear MDE's. ADecM is known as a reliable mathematical tool
for solving algebraic, differential or integro-differential linear and nonlinear
equations, but no general treatment for the type of nonlinear MDE's defined here
is available. ADecM has undergone several modifications to extend its field of
applicability, make it more efficient and speed up the convergence of the series
solution~\cite{arch,waz,zh,Riccati,hoss,waz-say,CMES}. In our application we specialize in the case of nonlinear MDE's describing stationary cylindrically symmetric (SCS) metrics (with four Killing vectors) in 5DGR extended by a Gauss-Bonnet (GB) term~\cite{Azreg,AzregN} and investigate them systematically. Knowing that no available technique of integration of differential equations can ever systematically handle this type of MDE's, our approach provides a systematic method for tackling them with the hope that it might be applied to MDE's in $d$DGR.

Some exact solutions to these MDE's describing SCS configurations in
5DGR have been constructed either by a perturbation approach~\cite{Azreg} or analytically~\cite{AzregN} and interpreted as cosmic strings. Since the MDE's are \textit{nonlinear}, they may admit different solutions and the quest for further physically interesting exact solutions is still an open question, which may lead to construct cosmic strings endowed with similar geometric properties as the 4DGR ones. In fact, the 5DGR superconducting cosmic strings constructed so far~\cite{Azreg,AzregN} exhibit different properties than their homologs in 4DGR~\cite{super}. Besides, there has been recently a renewed interest in generalized \`{a} la GB GR theories~\cite{024002,Thi,024018,044007,084025,024042,086002,104003,123515,Berej} and particularly in string theory fueled by the discovery of Capodimonte-Sternberg-Lens Candidate, CSL-1, a pair of aligned galaxies which lie at a redshift of $z = 0.46$ whose double image could be a result of gravitational lensing caused by
a cosmic string.

At the time Ref~\cite{Azreg} was written ADecM was being developed and not known to many workers (rather, it is still not known to many others), furthermore no other reliable method was available to tackle this type of MDE's, this is why we resorted to a perturbation approach. On the other hand, the alternative analytical method developed in~\cite{AzregN} is not suitable for a systematic investigation of this type of MDE's. Thus, our aim is to further develop ADecM to work \textit{systematically} for this type of MDE's in order to exhaust their set of possible solutions; this will turn out to be true for the case of the field equations describing SCS metrics in 5DGR extended by a GB term (sections~\ref{nts} \& \ref{ts}).

In section~\ref{for} we outline the properties of the MDE's~\eqref{1} (or \eqref{3}) and in section~\ref{sfield} we describe the physical problem and introduce the field equations---the Einstein-Gauss-Bonnet equations in the SCS case, which are of the form~\eqref{1} with $N=4$. In section~\ref{est} we develop and apply ADecM to the nonlinear field equations defined in section~\ref{sfield} and prove some of their properties. The most general unique power series solutions are constructed in sections~\ref{nts} \& \ref{ts} then reduced to closed forms.  These \textit{re-derived} exact solutions are shown to exhaust the set of possible solutions to the field equations describing single SCS configurations in 5DGR extended by a GB term. Following the conclusion section, Appendix A completes the proof of corollary 3. and Appendix B that of Corollary 6.

\section{Properties of the matrix differential equations}\label{for}
The formally defined nonlinear MDE~\eqref{3}
or~\eqref{1} has the following properties
\begin{enumerate}
  \item Since the invariants of a matrix $\chi$ remain invariant under a similarity
  transformation $S$, equation~\eqref{3} remains invariant if $\chi$ is replaced by
  \begin{equation}\label{pro1}
    \chi'= S\chi S^{-1}\,.
  \end{equation}
  Since the rank of a matrix is also invariant under a similarity
  transformation~\cite{Ox}, one can still generalize equation~\eqref{3} by making the
  functions $(G_j,R_j)$, $0\leq j\leq N-1$, depend explicitly on the rank of $\chi$
  without affecting this property;
  \item equation~\eqref{3} remains invariant under the simultaneous sign inversions $\chi\rightarrow -\chi$ \& $\rho\rightarrow -\rho$;
  \item It can be rigorously shown that any solution $\chi(\rho)$
  to equation~\eqref{3} commutes with its derivative $\chi_{,\rho}$~\cite{Sant}:
\begin{equation}\label{pro2}
    [\chi, \chi_{,\rho}]\equiv 0\,;
\end{equation}
  \item Relying on~\eqref{pro2} one can also show that any solution $\chi(\rho)$
  to equation~\eqref{3} is a polynomial in a constant matrix $A$ with scalar coefficients
  \begin{equation}\label{pro3}
    \chi(\rho)=\sum_{i=0}^{N-1}\alpha_i(\rho)A^i\,.
\end{equation}
\end{enumerate}
Since Property 3. results from Property 4., the two properties are equivalent.
It is also possible to first prove equation~\eqref{pro3} using a decomposition method, as this is done in section~\ref{est} for the field equations describing line sources in 5DGR.

\section{The field equations describing 5-dimensional cosmic strings}\label{sfield}
In 5DGR or Kaluza-Klein Theory (KKT) the spacetime has the topology
of the product $V^4\times S^1$, where $V^4$ is topologically equivalent to a
4-dimensional minkowskian spacetime and $S^1$ is a circle of radius $a$ parameterized by
the fifth coordinate $x^5$: $0\leq x^5\leq 2\pi a$. The fifth coordinate being spacelike
and periodic, the spacetime has then one timelike coordinate and four spacelike coordinates.

In 5DGR it is possible to modify the field equations by adding the GB term to the Lagrangian in such a way that the generalized Einstein-Gauss-Bonnet equations still involve at most second order derivatives of the metric~\cite{Lovelock}. The generalized action~\cite{Madore85} reads upon ignoring a cosmological term, which is not relevant for the purpose of constructing cosmic string solutions
\begin{equation*}
    S = -\frac{1}{16\,\pi\,G_{5}}\,\int {\rm d}^5x\,\sqrt{g}\,
\left(R + \frac{\gamma}{2}\, {\cal L}\right),
\end{equation*}
where $\gamma$ is a constant and the second term in the action is
the quadratic GB contribution ${\cal L} \equiv R^{ABCD}R_{ABCD} - 4\,R^{AB}R_{AB} + R^{2}$
(the upper-case Roman indices take the values 1 to 5).
The field equations derived upon varying the action with respect
to the metric components $g_{AB}$ are then
\begin{equation}\label{field}
    R_{AB} - (1/2)R\,g_{AB} + \gamma\,L_{AB} = 0,
\end{equation}
where $L_{AB}$ is the covariantly conserved ($L^{AB}_{\quad\,
;A}=0$) Lanczos tensor~\cite{Madore85}: $L_{AB} \equiv R_{A}{^{CDE}}R_{BCDE} - 2\,R^{CD}R_{ACBD} -
2\,R_{AC}R_{B}{^{C}} + RR_{AB} - (1/4)g_{AB}\,{\cal L}$.

The 5-metric of the 5-spacetime reads
\begin{equation}\label{met1}
    \text{d}s^{2} = g_{AB}\,
    \text{d}x^{A}\,\text{d}x^{B}\,,\;(A,B: 1 \rightarrow 5)\,.
\end{equation}
We restrict ourselves to a SCS 5-metric~\eqref{met1}, which has four commuting Killing vectors ($\partial_\varphi,\partial_z,\partial_t,\partial_5$), $\partial_t$ is timelike, and ($\partial_\varphi,\partial_5$) have closed orbits. Such a metric can be brought to the following form in an adapted coordinate system~\cite{Azreg}
\begin{equation}\label{met2}
    \text{d}s^{2} = -\,\text{d}{\rho}^{2} + {\lambda}_{ab}(\rho)\,
    \text{d}x^{a}\,\text{d}x^{b}\,,\;(a,b: 2 \rightarrow 5)\,,
\end{equation}
(the lower-case Roman indices take the values 2 to 5), where $x^1=\rho$ is a
radial coordinate, $x^2=\varphi$ (periodic) \& $x^3=z$ are the two other cylindrical
coordinates, $x^4=t$ (time), $g_{11}=-1$, $g_{1a}\equiv 0$ and $g_{ab}\equiv \lambda_{ab}(\rho)$ is
a 4$\times$4 real symmetrical matrix of signature (-- -- + --).

In GR the metric $g_{AB}$, in our case $\lambda$, is the main variable, however, in this paper, since we deal with the mathematics of the problem, we focus on the 4$\times$4 real matrix $\chi \equiv {\lambda}^{-1}{\lambda}_{,\rho}$ which we consider as the main variable. In terms of $\chi$ the field equations of the pure KKT corresponding to the metric~\eqref{met2} read~\cite{Azreg,AzregN}
\begin{eqnarray}
\label{pkkm} & &2\chi_{,\rho}+4\tr \chi_{,\rho}+(\tr \chi)\chi+ \tr \chi^{2}+(\tr \chi)^{2}=0\,, \\
\label{pkks} & &6\tr B+(\tr \chi)^{2}-\tr \chi^{2}=0 \,,
\end{eqnarray}
where $B \equiv \chi_{,\rho} + (1/2)\chi^{2}$. Notice that~\eqref{pkkm} is of the form~\eqref{1} and~\eqref{pkks} is an extra constraint on the invariants of $\chi$.

It is straightforward to bring equation~\eqref{pkkm} to the form~\eqref{3}. Subtracting the trace of equation~\eqref{pkkm} from equation~\eqref{pkks} $\times$ 3 we obtain the equation $\tr \chi^{2}=(\tr \chi)^{2}$ which we substitute into equation~\eqref{pkkm} to arrive at
\begin{equation}\label{pkkm1}
    2(4\tr \chi+2\chi)_{,\rho}+\tr \chi (4\tr \chi+2\chi)=0\,,\;\text{with}\;\tr \chi^{2}=(\tr \chi)^{2}\,.
\end{equation}
The system of equations~\eqref{pkkm} \& \eqref{pkks}, or its equivalent form~\eqref{pkkm1}, has been exactly solved in~\cite{Azreg}.

For a SCS 5-metric, the generalized Einstein-Gauss-Bonnet equations read~\cite{Azreg,AzregN}
\begin{eqnarray}
\label{m2} & &2\chi_{,\rho}+4\tr \chi_{,\rho}+(\tr \chi)\chi+ \tr \chi^{2}+(\tr \chi)^{2}+\gamma
\left\{(\chi^{3})_{,\rho}
-(\tr \chi)(\chi^{2})_{,\rho}\right. \nonumber \\
& &+\,[(\tr \chi)^{2}-\tr \chi^{2}]\chi_{,\rho} -(\tr \chi_{,\rho})[\chi^{2}-(\tr \chi)\chi]-(1/2)
(\tr \chi^{2})_{,\rho}\chi \nonumber \\
& &\left.+\,(1/2)[(\tr \chi)\chi^{3}-(\tr \chi^{2})\chi^{2} -(\tr \chi)(\tr \chi^{2})\chi+(\tr \chi)^{3}\chi]\right\}=0\,, \\
\label{s2} & &6\tr B+(\tr \chi)^{2}-\tr \chi^{2}+\gamma
\left\{\tr (B\chi^{2})-\tr (B\,\chi)\tr \chi \right. \nonumber \\
& &\left.+\,(1/2)\tr B[(\tr \chi)^{2}-\tr \chi^{2}]\right\}=0 \,,
\end{eqnarray}
where the extra terms with respect to equations~\eqref{pkkm} \& \eqref{pkks} are due to the Lanczos term. Equation~\eqref{s2} is an extra constraint on the invariants of $\chi$. As we mentioned in the Introduction, some \textit{exact} solutions to the system of equations~\eqref{m2} \& \eqref{s2} have been constructed either by a perturbation approach or analytically~\cite{Azreg,AzregN} without, however, exhausting the set of possible solutions. As application of our developed ADecM we will investigate systematically the system of equations~\eqref{m2} \& \eqref{s2}. Based mainly on ADecM and other techniques (differential equations and matrix theory), our approach is cognitive leading to the \textit{unique} exact solutions in closed forms, thus exhausting the set of all possible solutions to the system of equations~\eqref{m2} \& \eqref{s2}.

The system~\eqref{m2} \& \eqref{s2} is not readily solvable and need to be transformed to a reduced system as follows. Let the invariants of $\chi$ be the functions $f(\rho)\equiv \tr \chi$,
$g(\rho)\equiv \tr \chi^{2}$, $h(\rho)\equiv \tr \chi^{3}$ and $k(\rho)
\equiv \det \chi$, which have been called $\ell_1, \ell_2,\ell_3\;\text{and}\; k$ in equation~\eqref{1}. Let $G(\rho)\equiv
g-f^2$ and $H(\rho)\equiv h-f^3$. The Cayley-Hamilton equation for $\chi$ reads then
\begin{equation}\label{ch}
    \chi^{4}=f\chi^{3}+(G/2)\chi^{2}+[(H/3)-(fG/2)]\chi-k\,.
\end{equation}

We make use of the following
relations which are valid whether the matrices $\chi$ \&
$\chi_{,\rho}$ commute or not: $\tr (\chi_{,\rho}\chi)=g_{,\rho}/2$ \& $\tr (\chi_{,\rho}\chi^{2})=h_{,\rho}/3$. These relations together
with the trace of \eqref{ch} reduce \eqref{s2} to
\begin{eqnarray}
\label{in1} & &6[f_{,\rho}+(g/2)]-G+\gamma \left\{(h_{,\rho}/3)
+(fh/6)-(fg_{,\rho}/2) \right. \nonumber \\
& &\left.-\,(f^{2}g/2)+(g^{2}/4)+(f^{4}/12)-2k-
(G/2)[f_{,\rho}+(g/2)]\right\}=0\,.
\end{eqnarray}
Now, the trace of~\eqref{m2},
as provided by~\eqref{TrE}, also includes a term proportional to
($2h_{,\rho}+fh$) as equation~\eqref{in1} does
\begin{equation}\label{TrE}
  6f_{,\rho}+10f^2+8g+\gamma[2h_{,\rho}+fh-3fg_{,\rho}
   -4gf_{,\rho}+4f^2f_{,\rho}-f^2g+f^4-g^2]=0\,.
\end{equation}
Hence, eliminating ($2h_{,\rho}+fh$) by combining~\eqref{in1} and~\eqref{TrE},
we obtain $k$ in terms of $f$ \& $g$: $k=G(8+2\gamma G+3\gamma f^{2}+
 2\gamma f_{,\rho})$. Consequently, any solution to the system~(\ref{m2} \&
\ref{s2}) is necessary a solution to the following reduced system where $Q\equiv 4f+(2-\gamma G)\chi -
    \gamma f\chi^{2}+\gamma \chi^{3}$
\begin{eqnarray}
 \label{m3}  & & 2Q_{,\rho}+fQ+2G+\gamma G_{,\rho}\chi
 -\gamma G\chi^{2}=0\,, \\
 \label{s3}  & & 24\gamma k=G(8+2\gamma G+3\gamma f^{2}+
 2\gamma f_{,\rho})\,.
\end{eqnarray}
Equation~(\ref{m3}) has been readily obtained from~(\ref{m2}). In the case of pure KKT where the constant $\gamma =0$ the system~(\ref{m3}
\& \ref{s3}) leads to~\eqref{pkkm1}.

\section{The modified Adomian method}\label{est}

\subsection{Adomian's method}
ADecM has been developed~\cite{ad,Rach} to deal with the initial and/or boundary value problems in nonlinear science, which can be modeled by
\begin{equation}\label{p1}
    L_{\text{op}}u + R_{\text{op}}u + N_{\text{op}}u = s\,,
\end{equation}
where the unknown (scalar, vector or matrix) function $u(x)$ is subject to some initial and/or boundary conditions. The function $s(x)$ is a source term, ($L_{\text{op}}+R_{\text{op}}$) is a linear operator and $N_{\text{op}}$ includes nonlinear operator terms. The method consists in splitting the linear operator ($L_{\text{op}}+R_{\text{op}}$) into two linear terms, $L_{\text{op}}$ and $R_{\text{op}}$, where $L_{\text{op}}$ represents the highest order derivative and is easily invertible and $R_{\text{op}}$ groups the remaining lower order derivatives.

Applying the inverse operator $L_{\text{op}}^{-1}$ to both sides of~\eqref{p1} leads to a Volterra integral equation
\begin{equation*}
    u = w - L_{\text{op}}^{-1}(R_{\text{op}}u) - L_{\text{op}}^{-1}(N_{\text{op}}u)\,,
\end{equation*}
where $w$ is the sum of $L_{\text{op}}^{-1}(s)$ and the terms arising from the application of the initial and/or boundary conditions to $u$. If one decomposes $u$ into sum of components, $u = \sum_{m=0}^\infty u_m\,$, this leads to
\begin{equation}\label{p3}
    \sum_{m=0}^\infty u_m = w - L_{\text{op}}^{-1}\bigg(R_{\text{op}}\sum_{m=0}^\infty u_m\bigg) - L_{\text{op}}^{-1}\bigg(N_{\text{op}}\sum_{m=0}^\infty u_m\bigg)\,.
\end{equation}
Eq~\eqref{p1} is solved upon fixing recursively the components $u_m$, which can be done by first splitting the action of $N_{\text{op}}$ on $\sum_{m=0}^\infty u_m$ in~\eqref{p3} into sum of terms $\sum_{m=0}^\infty A^{\text{Ad}}_m$, where $A^{\text{Ad}}_m$ are called Adomian polynomials, in such a way that the first term $A^{\text{Ad}}_0$ depends only on $u_0$, which will be the first component of $u$ to be fixed, and $A^{\text{Ad}}_1$ depends only on ($u_0,u_1$), ..., and $A^{\text{Ad}}_m$ depends only on ($u_0,u_1,\dotsc, u_m$). Eq~\eqref{p3} takes then the form
\begin{equation}\label{p4}
    \sum_{m=0}^\infty u_m = w - L_{\text{op}}^{-1}\bigg(\sum_{m=0}^\infty R_{\text{op}}u_m\bigg) - L_{\text{op}}^{-1}\bigg(\sum_{m=0}^\infty A^{\text{Ad}}_m\bigg)\,.
\end{equation}

There is no unequivocal way to fix $u_0$~\cite{waz-say}, however, a straightforward way to do it is to identify $u_0(x)$ with $w(x)$ in~\eqref{p4} and the other components are determined by
\begin{equation*}
    u_{m+1} = - L_{\text{op}}^{-1}(R_{\text{op}}u_m) - L_{\text{op}}^{-1}(A^{\text{Ad}}_m),\;m\geq 0,\;\text{with}\;u_0=w\,,
\end{equation*}
where $A^{\text{Ad}}_m$ are evaluated using available formulas~\cite{ad,Rach,CMES}.

\subsection{The modified method}
Without loss of generality, we consider equation~(\ref{m3}) and develop our method by extending ADecM to deal with this type of equations; the approach readily extends to the general case~(\ref{3}). We split $\chi$ into sum of components
\begin{equation}\label{d1}
    \chi = \sum_{n=0}^\infty \,\chi_n \,,
\end{equation}
resulting in
\begin{equation}\label{d2}
    f = \sum_{n=0}^\infty f_n \,,\;\text{with}\;f_n = \tr \chi_n \,,
\end{equation}
and the linear operator $L_{\text{op}}$ according to ADecM should be $L_{\text{op}}=\text{d}/\text{d}\rho\,$. However, it is sometimes fruitful and time-saving to adjust $L_{\text{op}}$ to the kind of problem one is dealing with~\cite{arch}. For the case of equation~(\ref{m3}) our method consists to choose $L_{\text{op}}=\text{d}/\text{d}\rho + (f_0/2)$ the inverse operator of which reads
\begin{equation}\label{d3}
    L_{\text{op}}^{-1}[M]=\exp(-F_0(\rho))\int^{\rho}\exp(F_0(\rho'))M(\rho')\text{d}\rho'\,,
\end{equation}
where $M(\rho)$ is an arbitrary matrix and
\begin{equation}\label{d4}
    F_0(\rho)=
    (1/2)\int^{\rho} f_0(\rho')\text{d}\rho'\,.
\end{equation}
Since we do not assume initial or boundary conditions, the action of $L_{\text{op}}^{-1}$ on the zero matrix, and \textit{only} the zero matrix which appears in the r.h.s of~(\ref{m3}), results in
\begin{equation}\label{d5}
    L_{\text{op}}^{-1}[0]=\exp(-F_0(\rho))C\,,
\end{equation}
where $C$ is any constant matrix of integration. Replacing $f$ by its expansion~\eqref{d2} in equation~(\ref{m3}) then applying $L_{\text{op}}^{-1}$ to the latter and using the fact that $L_{\text{op}}^{-1}[2Q_{,\rho}+f_0Q]=2Q$, we obtain
\begin{equation}\label{r1}
    2Q=\exp(-F_0)C - L_{\text{op}}^{-1}[\underbrace{(f-f_0)}_{\sum_{n=1}^\infty f_n}Q+2G+\gamma G_{,\rho}\chi
 -\gamma G\chi^{2}]\,.
\end{equation}
Next, we choose to split  $Q$ in the l.h.s of~\eqref{r1} into a linear part $4f+2\chi$ and a nonlinear one $Q_{\text{non}}=-\gamma G\chi -\gamma f\chi^{2}+\gamma \chi^{3}\/$:
\begin{equation}\label{r2}
    Q=4f+2\chi +Q_{\text{non}}\,,
\end{equation}
and keep only the linear part on the l.h.s of~\eqref{r1}. Now, we apply the standard approach of ADecM and replace all the nonlinear contributions in~\eqref{r1}, that is
\begin{equation}\label{no}
    Q_{\text{non}}\quad \text{and}\quad \bigg(\sum_{n=1}^\infty f_n\bigg)Q+2G+\gamma G_{,\rho}\chi
 -\gamma G\chi^{2}
\end{equation}
by sums of components
\begin{align}
 \label{r3} & Q_{\text{non}}  = \sum_{n=0}^\infty Q_n(\chi_0,\dotsc,\chi_n)\,,\\
 \label{r3r} & \bigg(\sum_{n=1}^\infty f_n\bigg)Q+2G+\gamma G_{,\rho}\chi
 -\gamma G\chi^{2}=\sum_{n=0}^\infty B_n(\chi_0,\dotsc,\chi_n)\,,
\end{align}
in such a way that the components $Q_n$ and $B_n$ depend only on $(\chi_0,\dotsc,\chi_n)$ and do not depend on $\chi_m$ for $m>n$. Our method consists in defining these components such that if $N_{\text{op}}$ denotes any nonlinear operator, say, one of the operators in~\eqref{no} then $Q_0$ (resp. $B_0$) $=N_{\text{op}}[\chi_0]$, $Q_1$ (resp. $B_1$) $=N_{\text{op}}[\chi_0+\chi_1]-N_{\text{op}}[\chi_0]$, ..., $Q_n$ (resp. $B_n$) $=N_{\text{op}}[\chi_0+\dotsb +\chi_n]-N_{\text{op}}[\chi_0+\dotsb +\chi_{n-1}]$. Hence, the components $Q_n$ and $B_n$ are different from Adomian polynomials. For instance, the zeroth components are
\begin{align}
\label{ist1} Q_0(\chi_0)& =-\gamma G_0\chi_0 -\gamma f_0\chi_0^{2}+\gamma \chi_0^{3}\,,\\
\label{ist2} B_0(\chi_0)& =2 G_0 + \gamma G_{0,\rho}\chi_0 -\gamma G_0\chi_0^{2}\,,
\end{align}
where $G_0=\tr \chi_0^2-(\tr \chi_0)^2$. With that said, substituting Eqs~\eqref{d1}, \eqref{d2}, \eqref{r2}, \eqref{r3} \& \eqref{r3r} into~\eqref{r1} we bring it to
\begin{equation*}
    8f_0 + 4\chi_0 + \sum_{n=1}^\infty (8f_n + 4\chi_n) = \exp(-F_0)C - \sum_{n=0}^\infty \bigg(2Q_n + L_{\text{op}}^{-1}[B_n]\bigg)\,,
\end{equation*}
where the components $8f_n + 4\chi_n$, for $n\geq 0$, are determined recursively by\footnote{There is no unequivocal way to fix the zeroth component~\cite{waz-say} nor an unequivocal way to fix the other components. So in~\eqref{r5}, we could also choose $8f_0 + 4\chi_0 = \exp(-F_0)C -2Q_0$ without, however, affecting the following discussion.}
\begin{align}
\label{r5} 8f_0 + 4\chi_0 & = \exp(-F_0)C\,,\\
\label{r6} 8f_{n+1} + 4\chi_{n+1} & = -2Q_n - L_{\text{op}}^{-1}[B_n]\,,\;n\geq 0\,.
\end{align}

In order to derive an expression for $\chi_{n+1}$, for $n\geq 0$, we first trace~\eqref{r6} and use the fact that $\tr f_{n+1}=4f_{n+1}$ (recall we omit to write the identity matrix $I$) to obtain the formula
\begin{equation}\label{r6t}
    f_{n+1}=-\{2\tr Q_n + \tr (L_{\text{op}}^{-1}[B_n])\}/36\,,
\end{equation}
which we substitute back into~\eqref{r6} leading to
\begin{equation}\label{r6at}
    \chi_{n+1} = -\frac{2Q_n + L_{\text{op}}^{-1}[B_n]}{4} + \frac{2\tr Q_n + \tr (L_{\text{op}}^{-1}[B_n])}{18}\,.
\end{equation}

In the following two sections we shall discuss separately the cases $f\neq 0$ \& $f=0$.

\section{Nontraceless solutions: $\pmb{f\neq 0}$}\label{nts}
First we focus on the case $f\neq 0$. Without loss of generality, we can always
assume $f_0\neq 0$; if this were not the case we could permute the indices of the components $f_n$ and those of $\chi_n$ so that to make $f_0\neq 0$. The component $\chi_0$ is derived in the same way as we did with $\chi_{n+1}$: tracing~\eqref{r5} we obtain
\begin{equation}\label{r7}
    36f_0 = \exp(-F_0)\tr C \,,
\end{equation}
with $\tr C \neq 0$ since $f_0\neq 0$. We set $C=8(A+2)$ then differentiate~\eqref{r7} using~\eqref{d4} to obtain
\begin{align*}
    36f_{0,\rho} & = -(f_0/2)\exp(-F_0)\overbrace{8(\tr A +8)}^{\tr C}\\
    \quad &        = -(f_0/2)\times 36f_0
\end{align*}
leading to $f_{0,\rho}+f_0^2/2=0$ the solution of which, introducing a constant of integration $\rho_0$, reads
\begin{equation}\label{r8}
    f_0 = 2/(\rho - \rho_0)\,.
\end{equation}
Substituting~\eqref{r8} into~\eqref{d4}, we obtain
\begin{equation}\label{r9}
    \exp(F_0)=\rho - \rho_0\,.
\end{equation}
With $f_0$ and $\exp(F_0)$ given by~\eqref{r8} \& \eqref{r9}, equation~\eqref{r7} reduces to $\tr A =1$ and equation~\eqref{r5} to
\begin{equation}\label{r10}
    \chi_0 = 2A/(\rho - \rho_0)\,,\;\text{with}\;\tr A =1\,.
\end{equation}
Knowing $\chi_0\;\&\;f_0$ we derive the formula of the operator $L_{\text{op}}^{-1}$ using equation~\eqref{d3}
\begin{equation}
\label{r12} L_{\text{op}}^{-1}[M] = \frac{1}{\rho - \rho_0}\int^{\rho}(\rho' - \rho_0)M(\rho')\text{d}\rho' \,.
\end{equation}

Next we determine the first component $\chi_1$. With $\chi_0$ given by~\eqref{r10} we obtain $G_0 = 4(\tr A^2 - 1)/(\rho - \rho_0)^2$ which leads, using~\eqref{ist1} \& \eqref{ist2}, to
\begin{align*}
    Q_0 & = 8\gamma \,\frac{(A^3-A^2) - (\tr A^2 - 1)A}{(\rho - \rho_0)^3}\,,\\
    B_0 & = \frac{8(\tr A^2 - 1)}{(\rho - \rho_0)^2} - 16\gamma\,\frac{(\tr A^2 - 1)A + (\tr A^2 - 1)A^2}{(\rho - \rho_0)^4}\,.
\end{align*}
With $L_{\text{op}}^{-1}$ given by~\eqref{r12}, we obtain using~\eqref{r6at}
\begin{align}
\chi_1 & =  \frac{4\gamma [2(\tr A^3-\tr A^2)+(\tr A^2 - 1)^2]}{9(\rho - \rho_0)^3} -
    \frac{2(\tr A^2 - 1)\ln |\rho - \rho_0|}{9(\rho - \rho_0)}\nonumber \\
\label{k1} & \quad -\frac{4\gamma (A^3-A^2)+2\gamma (\tr A^2 - 1)(A^2 - A)}{(\rho - \rho_0)^3}\,.
\end{align}

Notice that $\chi_0$ \& $\chi_1$ are already of the form~\eqref{pro3} and all the other components are of the same form~\eqref{pro3}. Thus on its \textit{domain of convergence\,}, the matrix $\chi=\sum_{n=0}^\infty \chi_n$ too is of the form~\eqref{pro3}. In fact, if we try to obtain $\chi_2$ we first derive $Q_1$ \& $B_1$ using the formulas
\begin{align}
\label{ist3} & Q_1(\chi_0,\chi_1) =-\gamma (G_0+G_1)(\chi_0+\chi_1) -\gamma (f_0+f_1)(\chi_0+\chi_1)^{2}+\gamma (\chi_0+\chi_1)^{3} -Q_0(\chi_0)\,,\\
\label{ist4} & B_1(\chi_0,\chi_1) =2 (G_0+G_1) + \gamma (G_0+G_1)_{,\rho}(\chi_0+\chi_1) -\gamma (G_0+G_1)(\chi_0+\chi_1)^{2} -B_0(\chi_0)\,,
\end{align}
where $G_1=\tr (\chi_0+\chi_1)^2-(\tr \chi_0 + \tr \chi_1)^2 - G_0$ and $Q_0$ \& $B_0$ have been evaluated earlier. Since both matrices $\chi_0$ \& $\chi_1$ depend on the same matrix $A$ (they commute), the new matrices $Q_1$ \& $B_1$ too are functions of the same matrix $A$, consequently, $\chi_2$ given by~\eqref{r6at} in terms of $Q_1$ \& $B_1$ is also a function of $A$ which, after eliminating the powers of $A$ higher than 3 using the Cayley-Hamilton\footnote{The powers of $A$ higher than 3 result from the terms such as $\chi_0\chi_1$, $\chi_1^3$, etc in~\eqref{ist3} \& \eqref{ist4} and are eliminated using~\eqref{ch} where we have to replace $\chi$ and its invariants by $A$ and its invariants.} equation for $A$, reduces to the form~\eqref{pro3}. This argument applies to all remaining components. The whole argument remains valid and applies to the case $f\equiv 0$, too, as we will see in the next section.\\

\n \textit{Corollary 1.}\\
On its domain of convergence, the series representation, $\sum_{n=0}^\infty \chi_n$, of a solution to~\eqref{m3} converges to a function $\chi(\rho)$ of the form~\eqref{pro3}.\\

Notice that the constraint~\eqref{s3} has not been used yet. Our procedure consists first in determining the components $\chi_n$, as we have done in~\eqref{r10} \& \eqref{k1}, and once these are known we evaluate separately the invariants of $\chi$ ($k$, $f$ \& $G$) and substitute them into~\eqref{s3}, which turns into an \textit{algebraic} equation. From this point of view equation~\eqref{s3} splits as
\begin{align}
24\gamma \det (\chi_0+\chi_1+\dotsb) & =8(G_0+G_1+\dots) + 2\gamma (G_0+G_1+\dots)^2\nonumber\\
\quad & \quad + 3\gamma (G_0+G_1+\dots)(f_0+f_1+\dots)^{2}\nonumber\\
\label{s3n}\quad & \quad + 2\gamma (G_0+G_1+\dots)(f_0+f_1+\dots)_{,\rho}\,.
\end{align}
Using~\eqref{r10} \& \eqref{k1} it is straightforward to check that the r.h.s of~\eqref{s3n} includes a term proportional to $1/(\rho-\rho_0)^2$ emanating from $G_0 = 4(\tr A^2 - 1)/(\rho - \rho_0)^2$ while the l.h.s. does not. Consequently, in order to satisfy equation~\eqref{s3} we have to choose
\begin{equation*}
    G_0\equiv 0 \Rightarrow \tr A^2 = 1\,.
\end{equation*}

This new constraint on $\tr A^2$ reduces the expression of $\chi_1$, equation~\eqref{k1}, to
\begin{equation}\label{k1r}
    \chi_1 = \frac{8\gamma (\tr A^3-1)}{9(\rho - \rho_0)^3}
    -\frac{4\gamma (A^3-A^2)}{(\rho - \rho_0)^3}\,,
\end{equation}
where the terms proportional to $\ln (\rho-\rho_0)$ no longer appear and consequently will not appear in the remaining components ($\chi_m,\;m\geq 2$). Hence, each component $\chi_n$, $n\geq 0$, is a polynomial in $x=1/(\rho - \rho_0)$ with matrix coefficients, which are functions of the same constant matrix $A$, thus they are commuting matrices. Since the two nonlinear terms in~\eqref{no} remain \textit{invariant} under the simultaneous sign inversions $\chi\rightarrow -\chi$ \& $(\rho - \rho_0)\rightarrow -(\rho - \rho_0)$ (Property 2.) and since both $\chi_0$ and $\chi_1$ involve \textit{only odd powers} of $x$, by~\eqref{r6at} all the remaining components $\chi_m$, $m\geq 2$, will involve only odd powers with the lowest power being 3; $\chi_0$ is the only component involving a factor of $x$. If we group the same powers of $x^{2i+1}$, the general expression of $\chi$ solution to the system~\eqref{m3} \& \eqref{s3} is a power series in $x$. This rearrangement or grouping of the terms of the series $\sum_{n=0}^\infty \chi_n$ is justified as follows. Each element $\chi_{ab}$ ($a,b: 2 \rightarrow 5$) of the matrix series is a power series in $x$ which converges absolutely within some radius: $|x|<r_{ab}$. If $r_{\text{min}}$ denotes the minimum of all the radii $r_{ab}$, the series $\sum_{n=0}^\infty \chi_n$ converges \textit{absolutely} in the domain $|x|<r_{\text{min}}$ where the rearrangement of the terms is justified.\\

\n \textit{Corollary 2.}\\
Any nontraceless solution to the system~\eqref{m3} \& \eqref{s3} has two series representations, a) Adomian series $\sum_{n=0}^\infty \chi_n$ and b) a power series in $x^{2i+1}$ with constant commuting matrix coefficients $M_{2i+1}$, which are functions of the same constant matrix $A$ subject to $\tr A =\tr A^2 = 1$, of the form
\begin{equation}\label{kf}
    \chi = 2A\,x + \sum_{i=1}^\infty M_{2i+1}(A)\,x^{2i+1}\,,\;\tr A =\tr A^2 = 1\,,\;|x|<r_{\text{min}}\,,
\end{equation}
where $r_{\text{min}}$ is the radius of absolute convergence of both series representations.

By Corollary 2., the other invariants of $A$, that is, $\tr A^2$, $\det A$ \& $\text{rank}\,(A)$ are still free parameters. Corollary 3. puts more constraints on these remaining invariants.\\

\n \textit{Corollary 3.}\\
Any nontraceless solution to the system~\eqref{m3} \& \eqref{s3} has a power series representation~\eqref{kf} where the constant matrix $A$ is subject to
\begin{equation}\label{50b}
    \tr A =\tr A^2 = \tr A^3 = 1\;\& \;\det A = 0\,.
\end{equation}
A proof of Corollary 3. is detailed in Appendix A.

Now we proceed to fully determine the remaining components $\chi_n$ ($n\geq 2$) using the recursive formula~\eqref{r6}. Taking into account the constraints~\eqref{50b}, the expressions for $\chi_0$ \& $\chi_1$ (equations~\eqref{r10} \& \eqref{k1r}) simplify greatly and the Adomian series reduces to
\begin{equation}\label{f1}
    \chi = 2A\,x-4\gamma (A^3-A^2)\,x^3+\sum_{n=2}^{\infty}\chi_n\,,\;|x|<r_{\text{min}}\,.
\end{equation}
The constraints~\eqref{50b} reduce the Cayley-Hamilton equation for $A$ to $A^4=A^3$. Using this last equation in~\eqref{ist3} \& \eqref{ist4} leads to $Q_1=0$ \& $B_1=0$, which we substitute into~\eqref{r6} reads\footnote{We take $L_{\text{op}}^{-1}[B_1]=L_{\text{op}}^{-1}[0]=0$: only one global constant of integration $C$ is introduced in~\eqref{d5}.}
\begin{equation*}
    8f_2 + 4\chi_2 = 0\,\stackrel{\text{tracing}}{\Longrightarrow}\,f_2 =0\,\Rightarrow\,\chi_2 = 0\,.
\end{equation*}
With $\chi_2=0$ all the corresponding components are zero: $G_2=0$, $Q_2=0$ \& $B_2=0$ leading again to $f_3=0$ \& $\chi_3=0$ and so on $\chi_n=0$ for $n\geq 2$. Equation~\eqref{f1} reduces to the truncated solution with $\;r_{\text{min}}=\infty$
\begin{equation}\label{ff2}
    \chi = \frac{2A}{\rho - \rho_0}-\frac{4\gamma (A^3-A^2)}{(\rho - \rho_0)^3}\,,
\end{equation}
where $A$ is subject to~\eqref{50b}. It is straightforward to check that~\eqref{ff2} solves the system~\eqref{m3} \& \eqref{s3}.\\

\n \textit{Corollary 4.}\\
Any nontraceless solution to~\eqref{m3} \& \eqref{s3} is a singular matrix~\eqref{ff2} with $f=2/(\rho-\rho_0)$, $k=0$, $\tr A =\tr A^2 = \tr A^3 = 1$, $\det A = 0$ \& $r_{\text{min}}=\infty$.

The solution~\eqref{ff2} was derived earlier by a perturbation approach~\cite{Azreg} then analytically~\cite{AzregN}, but it has not been acknowledged in the literature that it is the unique nontraceless solution to~\eqref{m3} \& \eqref{s3}. Other locally equivalent as well as globally nonequivalent solutions to~\eqref{ff2} can be derived by similarity transformations~\eqref{pro1}~\cite{Azreg,AzregN}. A classification of the solutions~\eqref{ff2} according to $\text{rank}\,(A)$ has been discussed in~\cite{AzregN,Sant}.

In the standard Kaluza-Klein $4+1$ dimensional reduction, the components ($g_{15}/\lambda_{55}\equiv 0, \lambda_{25}/\lambda_{55},$ $\lambda_{35}/\lambda_{55}$, $\lambda_{45}/\lambda_{55}$) of the metric~\eqref{met2} are proportional the electromagnetic potentials ($A_1, A_2, A_3, A_4$), respectively, with $A_4$ being proportional to the electric potential. Rewrite Eq~(\ref{field}) as
\begin{equation}\label{field3}
    R^{A}{_{B}}-~(1/2)R\delta^{A}{_{B}}=8\pi GT_{\text{eff}}^{A}{_{\,B}}\,,
\end{equation}
where the r.h.s is the Lanczos term $-(\gamma/8\pi G)L^{A}{_{B}}$ reinterpreted as a 5-dimensional effective energy-momentum tensor source (here, $G$ is the gravitational constant). For the solution~\eqref{ff2} the evaluation the l.h.s of~\eqref{field3} depends explicitly on $\text{rank}\,(A)$. For $\text{rank}\,(A)=1\;\text{or}\;2$, the l.h.s of~\eqref{field3} includes only distributional contributions while for $\text{rank}\,(A)=3$ it includes distributional contributions as well as continuous ones, which are proportional to $\gamma$~\cite{Azreg,AzregN}. The solution~\eqref{ff2} with $\text{rank}\,(A)=1\;\text{or}\;2$ has been interpreted as neutral or charged cosmic string and that with $\text{rank}\,(A)=3$ as an extended superconducting cosmic string (the continuous contribution of $T_{\text{eff}}^{3}{_{\,5}}\to\infty$ as $\rho$$\to$0) surrounding a naked electrically charged cosmic string core in longitudinal translation carrying~\cite{AzregN} or not carrying~\cite{Azreg} an electric current.

\section{Traceless solutions: $\pmb{f = 0}$}\label{ts}
Next, we consider the other alternative: $f\equiv 0$. Notice that the above general formula, equation~\eqref{kf}, only applies to the case $f\neq 0$ since the trace of its r.h.s never vanishes \textit{identically} for any values of $\tr M_{2i+1}$ ($i\geq 1$). To analyze the case $f=0$ by the decomposition method we have to assume that at least one of the traces $f_n$ is zero, which we take it to be $f_0$. Hence, with $f_0 =0$, equation~\eqref{d4} leads to $F_0 =\text{constant}$ and the operator $L_{\text{op}}^{-1}$, equation~\eqref{d3}, reduces to
\begin{equation*}
\label{r13} L_{\text{op}}^{-1}[M] = \int^{\rho}M(\rho')\text{d}\rho' \,.
\end{equation*}
First we set\footnote{The matrix $A$ introduced for this case $f\equiv 0$ has nothing to do with previous notations.} $\exp(-F_0)C=4A$ in equation~\eqref{r5}, with $\tr A =0$ since $f_0 =0$, then apply the steps from~\eqref{r5} to~\eqref{r6at} to obtain $\chi_{n}$ ($n\geq 0$) as polynomials in $\rho$ with commuting matrix coefficients, which are functions of the same matrix $A$. The first two components rather written in the form~\eqref{pro3} read
\begin{align*}
 \chi_0  = A\,,\quad
 \chi_1  = \frac{2\gamma \tr A^3-9\tr A^2 + \tr A^2(8-\gamma \tr A^2)\rho}{18}
  +\frac{\gamma \tr A^2 (2+\rho)A - 2\gamma A^3}{4}\,.
\end{align*}
It is obvious that $\chi$ will have the form~\eqref{pro3}. Within its radius of absolute convergence, $r_{\text{min}}$, we can rearrange the terms of $\chi =\sum_{n=0}^\infty \chi_n$ by grouping the same powers of $\rho$, the matrix $\chi$ takes the following shape where $\tr M_i = 0$ ($i\geq 0$) to ensure that $f=\tr \chi \equiv 0$ and where $M_i(A)$ commute since they are functions of the same matrix $A$
\begin{equation}\label{kf0}
    \chi = \sum_{i=0}^\infty M_i \,\rho^i \;\text{for}\;\rho <r_{\text{min}}\,,\;\text{with}\;\tr M_i \equiv 0\;\text{for}\;i\geq 0\,.
\end{equation}

\n \textit{Corollary 5.}\\
Any traceless solution to the system~\eqref{m3} \& \eqref{s3} has two series representations, a) Adomian series $\sum_{n=0}^\infty \chi_n$ and b) a power series in $\rho^i$ with constant commuting matrix coefficients $M_i$, which are functions of the same constant matrix $A$ subject to $\tr A =0$, of the form~\eqref{kf0}.

Substituting the series expansion~\eqref{kf0} into equations~\eqref{m3}, \eqref{s3} \& \eqref{ch} we bring them to the following expansions
\begin{equation}\label{e0}
\eqref{m3}\Rightarrow \sum_{n=0}^\infty C_n\,\rho^n = 0\,,\;
\eqref{s3}\Rightarrow \sum_{n=0}^\infty c_n\,\rho^n = 0\,,\;
\eqref{ch}\Rightarrow \sum_{n=1}^\infty E_n\,\rho^n \equiv 0\,,
\end{equation}
where $C_n$ \& $E_n$ are matrix coefficients and $c_n$ are scalar ones depending on the traceless and commuting matrices $M_i$. Setting these matrix and scalar coefficients to zero will alow us to determine the matrices $M_i$ ($\tr M_i\equiv 0$ for $i\geq 0$).

In the calculation below we will make use of the Cayley-Hamilton equation for $M_0$ ($\tr M_0=0$)
\begin{equation}\label{ch2}
    M_0^4 = (\tr M_0^2/2)M_0^2 + (\tr M_0^3/3)M_0 - \det M_0\,.
\end{equation}

The equations $C_0=0$, $c_0=0$ \& $c_1=0$ lead respectively to
\begin{align}
\label{e1} & (4 -2 \gamma \tr M_0^2+6 \gamma  M_0^2)M_1+2 \tr M_0^2-\gamma  (\tr M_0^2) M_0^2 \nonumber \\
& -2 \gamma \tr (M_0M_1) M_0=0\,,\\
\label{e2} & \det M_0 = \tr M_0^2/(3\gamma)+(\tr M_0^2)^2/12\,,\\
\label{e3} & \gamma  \tr M_0^2 \tr (M_0M_1)-4 \tr (M_0M_1)-6 \gamma \tr (M_0^3M_1)=0\,.
\end{align}
Adding~\eqref{e3} to the trace of~\eqref{e1}$\times M_0$ leads to
\begin{equation}\label{e4}
    \tr M_0^2 [\tr M_0^3+3\tr (M_0M_1)]=0\,.
\end{equation}
First using the Cayley-Hamilton equation for $M_0$, equation~\eqref{ch2}, to eliminate any power of $M_0$ higher than 3 in the product~\eqref{e1}$\times M_0^2$ then combining the trace of~\eqref{e1} with that of~\eqref{e1}$\times M_0^2$ to eliminate $\tr (M_0^2M_1)$ leads to
\begin{equation*}
    \tr M_0^2 (\tr M_0^2-2/\gamma)=0\,.
\end{equation*}
From this last equation we have either 1) $\tr M_0^2=2/\gamma$ or 2) $\tr M_0^2=0$.

\subsection{$\pmb{\tr M_0^2=2/\gamma}$}
Equation~\eqref{e2} implies $\det M_0=1/\gamma^2$. Using the Cayley-Hamilton equation for $M_0$, equation~\eqref{ch2}, to eliminate any power of $M_0$ higher than 3 in the product~\eqref{e1}$\times M_0^3$, then combining the trace of~\eqref{e1}$\times M_0^3$ with~\eqref{e3} to eliminate $\tr (M_0^3M_1)$ we obtain $5\tr M_0^3+6\tr (M_0M_1)=0$. Solving this last equation together with equation~\eqref{e4}, which now reads $\tr M_0^3+3\tr (M_0M_1)=0$, results in $\tr M_0^3=0$ \& $\tr (M_0M_1)=0$. Substituting these values into~\eqref{e1} reads $6\gamma M_0^2M_1+(4/\gamma)-2M_0^2=0$, then multiplying it by $\gamma (1-\gamma M_0^2)$, which is the inverse matrix of $M_0^2$, leads to
\begin{equation}\label{e5}
    M_1 = -1/(3\gamma)+2M_0^2/3\,.
\end{equation}
With this value of $M_1$, equation~$E_1=0$ is identically satisfied.

Equations $C_1=0$ \& $c_2=0$ are brought to the following forms, respectively, using~\eqref{e5}
\begin{align}
\label{e6} & [2(\gamma -1)+3\gamma^3\tr (M_0M_2)]M_0 + 2\gamma^2 M_0^3 - 9\gamma^3 M_0^2M_2=0\,,\\
\label{e7} & 3 \gamma ^3 \tr (M_0M_2)+9 \gamma ^4 \tr (M_0^3M_2)-2-9 \gamma =0\,.
\end{align}
Solving~\eqref{e7} along with the trace of~\eqref{e6}$\times M_0$ we arrive at
\begin{equation}\label{aat}
    \tr (M_0M_2)=1/\gamma^3+3/(2\gamma^2)\,,\quad \tr (M_0^3M_2)=-1/(9\gamma^4)+1/(2\gamma^3)\,.
\end{equation}
Using the Cayley-Hamilton equation for $M_0$, equation~\eqref{ch2}, to eliminate any power of $M_0$ higher than 3 in the product~\eqref{e6}$\times M_0^3$, then eliminating $\tr (M_0M_2)$ \& $\tr (M_0^3M_2)$ by~\eqref{aat} we obtain $\gamma =2/3$. With this value of $\gamma$ and the relations~\eqref{aat}, the matrix equation $E_2=0$ reads
\begin{equation}\label{e8}
    (64/9)M_0^3M_2 - (16/3)M_0M_2 - (32/3)M_0^2 + 14 = 0\,.
\end{equation}
Tracing both sides of~\eqref{e8} and using~\eqref{aat} along with $\tr M_0^2=2/\gamma$ ($\gamma =2/3$), we obtain $-4=0$. Hence, the case $\tr M_0^2=2/\gamma$ does not lead to traceless solutions to the system~\eqref{m3} \& \eqref{s3}.

\subsection{$\pmb{\tr M_0^2=0}$}\label{2z}
Equation~\eqref{e2} implies $\det M_0=0$. In this case equation~\eqref{e4} is satisfied so that $\tr M_0^3$ remains undetermined. Now, if we combine the trace of~\eqref{e1}$\times M_0^3$ with~\eqref{e3} we obtain $\tr (M_0M_1)=0$, which reduces equation~\eqref{e3} to $\tr (M_0^3M_1)=0$. Substituting $\tr M_0^2=0$ \& $\tr (M_0M_1)=0$ into~\eqref{e1}, the remaining equation $(4 +6 \gamma  M_0^2)M_1=0$ leads upon multiplying by $(1/4)-3\gamma M_0^2/8$, which is the inverse matrix of $4 +6 \gamma  M_0^2$, to
\begin{equation*}
    M_1 = 0\,.
\end{equation*}

Hence, the traceless case ($\tr M_i\equiv 0$ for $i\geq 0$) with $\tr M_0^2=0$ leads to $\det M_0=0$, $M_1=0$ and $\tr M_0^3$ \& $M_0$ remain undetermined. As will be shown in Appendix B, even though $\tr M_0^3$ is still a free parameter, all the remaining matrices $M_i$, $i\geq 2$, are proven to be zero. We have thus obtained the following traceless and singular exact solution
\begin{equation}\label{ff3}
    \chi = M\,,\;\text{with}\;\tr M=\tr M^2=\det M=0\,,
\end{equation}
where $M$ is any $4\times 4$ constant matrix with $\tr M=\tr M^2=\det M=0$ and arbitrary $\tr M^3$ \& $\text{rank}\,(M)$.\\

\n \textit{Corollary 6.}\\
Any traceless solution to~\eqref{m3} \& \eqref{s3} is a singular constant matrix~\eqref{ff3} with $f=0$, $g=0$, $k=0$ and arbitrary $h$.

Looking for constant solutions to the system~\eqref{m3} \& \eqref{s3}, the solution~\eqref{ff3} was derived analytically in Ref.~\cite{Azreg}, but it has not been acknowledged in the literature that it is the unique traceless solution to~\eqref{m3} \& \eqref{s3}.

\section{Conclusion}\label{con}
Using ADecM we have shown, relying on the particular case~\eqref{m3}, that solutions $\chi(\rho)$ to equation~\eqref{3} are necessarily polynomials in a constant matrix with scalar coefficients, which are  functions of $\rho$ (equation~\eqref{pro3}). Restricting ourselves to the case of a 4$\times$4 nonlinear MDE along with a scalar one, which constitute the field equations of a 5DGR extended by a quadratic GB term in the SCS case (equations~\eqref{m3} \& \eqref{s3}), we have been able to construct the general \textit{unique power} series solutions (equations~\eqref{kf} \& \eqref{kf0}). These power series terminate in closed-form singular ($\det \chi =0$), nontraceless~\eqref{ff2} and traceless~\eqref{ff3} solutions, which all satisfy the property $\tr \chi^2-(\tr \chi)^2\equiv 0$. No nonsingular ($\det \chi \neq 0$) or further singular solutions have emerged from the unique power series solutions. This constitutes a proof of uniqueness of the re-derived solutions~\eqref{ff2} \& \eqref{ff3}.

In summary, we have developed in this paper an efficient ADecM to solve the constrained nonlinear MDE's of type~\eqref{m3}. The search for SCS solutions in $d$DGR ($d\geq 5$) leads to a system of $(d-1)\times(d-1)$ MDE along with a scalar one, solutions of which may be laid down by the ADecM developed here.

\section*{Acknowledgments}
The author wishes to thank G\'erard Cl\'ement (LAPTH) for useful comments on this manuscript.

\section*{Appendix A: Proof of Corollary 3.}
\appendix
\def\theequation{A.\arabic{equation}}
\setcounter{equation}{0}
In equation~\eqref{kf} we set $M_1=2A$. The constraints on $A$ become: $\tr M_1=2$
\& $\tr M_1^2=4$. Now, $\chi$ being provided by~\eqref{kf}, the general expressions up to the order 7 for $G$ and $k$ read
\begin{align*}
 G &= [\tr M_1^2-(\tr M_1)^2]x^2+[2 \tr (M_1M_3)-2
\tr M_1 \tr M_3]x^4 \\
 \quad & \quad +[2 \tr (M_1M_5)+\tr M_3^2-(\tr M_3)^2-2 \tr M_1 \tr M_5]x^6 + \mathcal{O}(x^8)\,,
\end{align*}
\begin{align*}
    k &=[(\tr M_1)^4-6 (\tr M_1)^2 \tr M_1^2+3 (\tr M_1^2)^2+8 \tr M_1 \tr M_1^3-6 \tr M_1^4]x^4/24 \\
& \quad +[2
(\tr M_1)^3 \tr M_3-6
\tr M_1 \tr M_3
\tr M_1^2-6 (\tr M_1)^2 \tr (M_1M_3) -12
\tr (M_1^3M_3) \\
 & \quad +6
\tr M_1^2 \tr (M_1M_3)+4 \tr M_3 \tr M_1^3+12 \tr M_1 \tr (M_1^2M_3)]x^6/12 + \mathcal{O}(x^8)\,,
\end{align*}
Substituting the above expressions and values into~\eqref{m3} \& \eqref{s3} and using the Cayley-Hamilton equation for $M_1$, the terms of order 4 in these equations lead to the
matrix \& scalar equations, respectively
\begin{align}
\label{q6} & 4\gamma (2M_1^2 - M_1^3) +
2(\tr M_3)\,M_1 - 8M_3  -16\tr M_3 +
4\tr (M_1M_3) = 0\,,\\
\label{q5} & 3\gamma \det M_1 + 4\tr M_3 - 2\tr (M_1M_3) = 0\,.
\end{align}
In general the terms of order $l+1$ ($l$ odd) in~\eqref{m3} \& \eqref{s3}
lead to matrix \& scalar equations, respectively, depending
linearly on $\tr M_l\; \& \;\tr (M_1M_l)$ as
in~\eqref{q6} \& \eqref{q5}. Furthermore, the matrix equation
depends linearly on $M_l$. Such a system can always be solved for
($\tr M_l$, $\tr (M_1M_l)$, $M_l$) as we are going to do
for~\eqref{q6} \& \eqref{q5}. Since the resolution of the system
involves tracing the matrix equation, consistency of the obtained
solution ($\tr M_l$, $\tr (M_1M_l)$, $M_l$) has to be
checked for each step.

Solving the system consisted of~\eqref{q5} and the trace
of~\eqref{q6}, we obtain
\begin{align}
\label{q7}  \tr M_3 & = \frac{2}{3} \gamma  \det M_1 + \frac{\gamma }{9} (8-\tr M_1^3)\\
\label{q8} \tr (M_1M_3) & = \frac{17}{6} \gamma \det M_1
+ \frac{2 \gamma }{9} (8-\tr M_1^3)\,.
\end{align}
Substituting these last two equations in~\eqref{q6}, we obtain
\begin{align}
  M_3 & = \frac{\gamma }{2} (2 M_1^2 - M_1^3) + \frac{1}{4} \bigg(\frac{2 \gamma }{3} \det M_1 + \frac{\gamma }{9}
    (8-\tr M_1^3)\bigg)M_1\nonumber \\
\label{q9} & \quad + \frac{\gamma }{12}\,\det M_1 - \frac{\gamma }{9}(8-\tr M_1^3)\,.
\end{align}
Tracing~\eqref{q9} reduces to~\eqref{q7} and
tracing~\eqref{q9}$\times M_1$ reduces to~\eqref{q8}. Hence, the
solution ($\tr M_3$, $\tr (M_1M_3)$, $M_3$) is
consistent.

We applied the same procedure for the order 6
and obtained the following solution ($\tr M_5$, $\tr (M_1M_5), M_5$)
\begin{align}
\tr M_5 & =\frac{\gamma ^2}{972}\,[1024+1698
\det M_1+171 (\det M_1)^2] \nonumber \\
\label{q13} & \quad -\frac{\gamma
\tr M_1^3}{486}\,(128 \gamma +96 \gamma
\det M_1) +\frac{4\gamma
^2 (\tr M_1^3)^2}{243}\,, \\
\tr (M_1M_5) & = \frac{\gamma ^2 [-512+13164
\det M_1+2223 (\det M_1)^2]}{1944}
\nonumber \\
\label{q14} & \quad -\frac{\gamma  (-128
\gamma +1929 \gamma  \det M_1)
\tr M_1^3}{1944}-\frac{\gamma ^2
(\tr M_1^3)^2}{243}\,,
\end{align}
\begin{align}
 M_5 & = \left(-\frac{\gamma ^2}{3}-\frac{\gamma ^2 \det M_1}{4}+\frac{\gamma ^2 \tr M_1^3}{24}\right) M_1^3\nonumber \\
& \quad + \left(\frac{2 \gamma ^2}{9}+\frac{17 \gamma ^2 \det M_1}{48}-\frac{\gamma ^2 \tr M_1^3}{36}\right) M_1^2\nonumber \\
& \quad +\left(\frac{254 \gamma ^2}{243}+\frac{547 \gamma ^2
\det M_1}{1296}+\frac{31\gamma ^2 (\det M_1)^2}{864}
-\frac{73 \gamma ^2
\tr M_1^3}{486}\right. \nonumber \\
\label{q15} & \quad \left. -\frac{19\gamma ^2 \det M_1
\tr M_1^3}{648} +\frac{19 \gamma ^2
(\tr M_1^3)^2}{7776}\right) M_1 +\frac{5\gamma ^2
(\det M_1)^2}{48}-\frac{37 \gamma ^2 \det M_1}{72}\nonumber \\
& \quad  +\frac{13 \gamma ^2 \tr M_1^3}{27} - \frac{52 \gamma
^2}{27}+\frac{\gamma
^2 \det M_1 \tr M_1^3}{9}
-\frac{13 \gamma ^2 (\tr M_1^3)^2}{432}\,.
\end{align}
The solution ($\tr M_5$, $\tr (M_1M_5)$, $M_5$) failed to
be consistent. In fact, subtracting~\eqref{q13} from the trace
of~\eqref{q15} and subtracting~\eqref{q14} from the trace
of~\eqref{q15}$\times M_1$ lead to the following two constraints
on the two free parameters ($\tr M_1^3$, $\det M_1$)
\begin{align}
\label{q16}& 45 \gamma  (\det M_1)^2-13 \gamma
(\tr M_1^3-8)^2 +6 \gamma \det M_1
 (8 \tr M_1^3-37) =0\,,\\
\label{q17}& 90 \gamma  (\det M_1)^2-8 \gamma
(\tr M_1^3-8)^2+3 \gamma \det M_1
 (89 \tr M_1^3-496) =0\,.
\end{align}

Substituting the series expansion of $\chi$ (equation~\eqref{kf})
into its Cayley-Hamilton equation~\eqref{ch} and replacing
the matrices $M_3,\,M_5$ and their traces by the expressions
derived above without, however, using the
constraints~\eqref{q16} \& \eqref{q17}, we obtain upon rearranging the terms the
following ``matrix'' series expansion
\begin{align}
\label{q21} & \{[(16 \gamma +6 \gamma  \det M_1+9 (\tr M_2)^2) \tr M_1^3-64 \gamma -48 \gamma  \det M_1-\gamma  (\tr M_1^3)^2]
M_1 \\
&  +96 \gamma  \det M_1 -128 \gamma +(32 \gamma -12 \gamma  \det M_1) \tr M_1^3-2 \gamma  (\tr M_1^3)^2 \}x^6/36 + \mathcal{O}(x^8) \equiv 0\,.\nonumber
\end{align}
Notice that the matrix coefficients of $x^4\;\& \;x^5$ in~\eqref{q21} are identically zero. Since~\eqref{q21} is an identity, all the matrix coefficients have to be zero. Hence, tracing the matrix coefficient of $x^6$ in~\eqref{q21}, we obtain the following constraint
\begin{equation}
\label{q22} [18 \gamma  \det M_1+5 \gamma  (\tr M_1^3-8)] (\tr M_1^3-8) = 0\,.
\end{equation}
Solving the system of the three constraints~\eqref{q16}, \eqref{q17} \& \eqref{q22} leads to
\begin{equation*}
 (\det M_1 = 0\; \&\;  \tr M_1^3 = 8) \Rightarrow  (\det A = 0\; \&\;  \tr A^3 = 1)\,.
\end{equation*}

\section*{Appendix B: Proof of Corollary 6.}
\appendix
\def\theequation{B.\arabic{equation}}
\setcounter{equation}{0}
Given $f=0$, $\tr M_0=0$ \& $\tr M_0^2=0$, we have shown in subsection~\ref{2z} that $\tr M_0^3$ \& $M_0$ remain arbitrary and that $\det M_0=0$ \& $M_1=0$. To show that the other matrices $M_j$, $j\geq 2$, are also zero we conduct a proof by induction. Knowing that $M_1\equiv 0$ we assume that, for $j\geq 2$, $M_1=M_2=\dotsb =M_{j-1}\equiv 0$ and will show that $M_j\equiv 0$. $\chi$, provided by~\eqref{kf0}, and its m\textsuperscript{th} power become
\begin{align*}
 & \chi = M_0 +M_j\,\rho^j +M_{j+1}\,\rho^{j+1}+\dotsb \,,\\
 & \chi^m = M_0^m +m\,M_0^{m-1}\,M_j\,\rho^j +\dotsb \,,
\end{align*}
With $M_1=M_2=\dotsb =M_{j-1}\equiv 0$ for $j\geq 2$, equations~\eqref{e0} become
\begin{equation*}
\hspace{-1.1mm}\eqref{m3}\Rightarrow \sum_{n=j-1}^\infty C_n\,\rho^n = 0,\,
\eqref{s3}\Rightarrow \sum_{n=j}^\infty c_n\,\rho^n = 0,\,
\eqref{ch}\Rightarrow \sum_{n=j}^\infty E_n\,\rho^n \equiv 0.
\end{equation*}
Since for $f=0$ we have $\tr M_j=0$, the scalar equation~$c_j =0$ leads immediately to $\tr (M_0M_j)=0$. Using this last equation in the matrix equation $C_{j-1} =0$ we reduce it to
\begin{equation}\label{ap4}
    (4 + 6\gamma M_0^2)M_j = 0\,.
\end{equation}
Multiplying both sides of~\eqref{ap4} by $(1/4)-3\gamma M_0^2/8$, which is the inverse matrix of $4 +6 \gamma  M_0^2$, leads to $M_j =0$. With this value of $M_j$, the matrix equation $E_j =0$ vanishes identically.


\begin{thebibliography}{99}
\bibitem{Campbell}S.L. Campbell, ``Singular Systems of
Differential Equations II," Pitman, Marshfield, MA (1982).
\bibitem{Lewis}F.L. Lewis,
Circ. Syst. Signal Process {\bf 5}(1) (1986) 3.
\bibitem{Azreg}M. Azreg-A\"{\i}nou and G. Cl\'ement G,
Class.
Quantum Grav. {\bf 13} (1996) 2635.
\bibitem{AzregN}M. Azreg-A\"{\i}nou,
 EPL
{\bf 81}(6) (2008) 60003.
\bibitem{Maison} D. Maison, Gen. Relativ. Gravit.
{\bf 10} (1979) 717;\\
P. Dobiasch and D. Maison, Gen. Relativ. Gravit. {\bf 14} (1982) 231.
\bibitem{AzregC} M. Azreg-A\"{\i}nou, G. Cl\'ement, C.P. Constantinidis and
J.C. Fabris, Grav. Cosmol. {\bf 6} (2000) 207
[gr-qc/9911107].
\bibitem{Clement}G. Cl\'ement and I. Zouzou, Phys. Rev. {\bf
D50} (1994) 7271.
\bibitem{Wild}W. Wild, [gr-qc/9812095] (1998).
\bibitem{Ox}P. Lancaster and M. Tismenetsky, ``The Theory of Matrices: with
Applications," Academic Press, Orlando (1985).
\bibitem{ad}G. Adomian, Solving Frontier Problems of Physics: The Decomposition Method, Kluwer, Boston, MA (1994).
\bibitem{Rach}R.C. Rach, Kybernetes {\bf 37} (2008) 910-955.
\bibitem{arch}P. Di\c{t}\u{a} and N. Grama, [solv-int/9705008v1] (1997);\\
    A.M. Wazwaz, Appl. Math. Comput. {\bf 173} (2006) 165-176.
\bibitem{waz}A.M. Wazwaz, Found. Phys. Lett. {\bf 13} (2000) 493-498.
\bibitem{zh}X. Zhang, J. Comput. Appl. Math. {\bf 180} (2005) 377-389.
\bibitem{Riccati}M.A. El-Tawil, A.A. Bahnasawi and A. Abdel-Naby, Appl. Math. Comput. {\bf 157}(2) (2004) 503-514.
\bibitem{hoss}M.M. Hosseini, Appl. Math. Comput. {\bf 181} (2006) 1737-1744.
\bibitem{waz-say}A.M. Wazwaz, S.M. El-Sayed, Appl. Math. Comput. {\bf 122} (2001) 393-405.
\bibitem{CMES}M. Azreg-A\"{\i}nou, CMES: Computer Modeling in Engineering \& Sciences {\bf 42} (2009) 1-18.
\bibitem{super}B. Linet, Class.
Quantum Grav. {\bf 6} (1989) 435-442.\\
T.M. Helliwell, Phys. Lett. A \textbf{143} (1990) 438-442.
\bibitem{024002}G. Dotti, J. Oliva and R. Troncoso, Phys. Rev. {\bf D 75} (2007) 024002.
\bibitem{Thi}M. Thibeault, C. Simeone and E.F. Eiroa, Gen. Relativ. Gravit. {\bf 38} (2006) 1593.
\bibitem{024018}K. Farakos and P. Pasipoularides, Phys. Rev. {\bf D 75} (2007) 024018.
\bibitem{044007}H. Maeda and N. Dadhich, Phys. Rev. {\bf D 75} (2007) 044007.
\bibitem{084025}E. Gravanis and S. Willison, Phys. Rev. {\bf D 75} (2007) 084025.
\bibitem{024042}M. Melis and S. Mignemi, Phys. Rev. {\bf D 75} (2007) 024042.
\bibitem{086002}G. Cognola, E. Elizalde, S. Nojiri, S.D. Odintsov and S. Zerbini, Phys. Rev. (D 75) (2007) 086002.
\bibitem{104003}K. Konya, Phys. Rev. {\bf D 75} (2007) 104003.
\bibitem{123515}J.D. Barrow and J. Middleton, Phys. Rev.
{\bf D 75} (2007) 123515.
\bibitem{Berej}W. Berej, J. Matyjasek, D. Tryniecki and M. Woronowicz, Gen. Relativ. Gravit. {\bf 38} (2006) 885.
\bibitem{Sant}M. Azreg-A\"{\i}nou, in ``Integral Methods in Science and Engineering, Analytic Methods," Volume 1, Chapter 5, a Birkh\"{a}user book (Boston), Editors: C. Constanda, M.E. P\'erez. (http://www.springer.com/birkhauser/mathematics/book/978-0-8176-4898-5).

\bibitem{Lovelock}D. Lovelock, J. Math. Phys. {\bf 12} (1971) 498-501.
\bibitem{Madore85}J. Madore, Phys. Lett. {\bf 110}A (1985) 289.

\end{thebibliography}
\end{document}